\documentclass[mathleft
]{an}
\usepackage{aas_macros}
\usepackage[sort,authoryear]{natbib}
\usepackage{graphicx}
\usepackage{times}
\usepackage{paralist}
\usepackage{verbatim}
\begin{document}
%
\Pagespan{489}{501}
\Yearpublication{2010}%
\Yearsubmission{2010}%
\Month{05}%
\Volume{331}%
\Issue{5}%
\DOI{10.1002/asna.201011366}%
\title{Pulsations and planets:\\ the asteroseismology-extrasolar-planet connection%
}
\author{S. Schuh\inst{1,2}\fnmsep
  \thanks{Corresponding author:
  \email{schuh@astro.physik.uni-goettingen.de}\newline}
}
\titlerunning{Pulsations and planets}
\authorrunning{S. Schuh}
\institute{
  TEA Visiting Professor, Eberhard-Karls-Universit\"at T\"ubingen, 
  Kepler Center for Astro and Particle Physics, 
  Institut f\"ur Astronomie und Astrophysik,
  Sand~1,
  72076 T\"ubingen, Germany
  \and
  Georg-August-Universit\"at G\"ottingen, Institut f\"ur Astrophysik,
  Friedrich-Hund-Platz~1, 37077 G\"ottingen, Germany
}
\received{2010}
\accepted{2010}
\publonline{05/2010}
\keywords{stars: evolution  --
  fundamental parameters --
  horizontal-branch --
  oscillations -- 
  planetary systems 
}
\abstract{%
The disciplines of asteroseismology and extrasolar planet science overlap
methodically in the branch of high-precision photometric time series
observations. Light curves are, amongst others, useful to measure intrinsic
stellar variability due to oscillations, as well as to discover and
characterize those extrasolar planets that transit in front of their host
stars, periodically causing shallow dips in the observed brightness. Both
fields ultimately derive fundamental parameters of stellar and planetary
objects, allowing to study for example the physics of various classes of
pulsating stars, or the variety of planetary systems, in the overall context
of stellar and planetary system formation and evolution. Both methods
typically also require extensive spectroscopic follow-up to fully explore the
dynamic characteristics of the processes under investigation. In particularly
interesting cases, a combination of observed pulsations and signatures of a
planet allows to characterize a system's components to a very high degree of
completeness by combining complementary information. The planning of the
relevant space missions has consequently converged with respect to science
cases, where at the outset there was primarily a coincidence in
instrumentation and techniques. Whether space- or ground-based, a specific
type of stellar pulsations can themselves be used in an innovative way to
search for extrasolar planets. Results from this additional method at the
interface of stellar pulsation studies and exoplanet hunts in a
beyond-mainstream area are presented.%
}
\maketitle
\section{Introduction}
\label{sec:introduction}
The (transiting) extrasolar planet fields and the asteroseismology
field see a convergence of instrumentation that culminates in the
insight that beyond this purely technical level, a much more
fundamental connection exists in the shared desire for the most
exhaustive characterization of stellar and planetary systems at all
possible with the available diagnostics.
I will first sum up the relevant current context with an emphasis on
planets around evolved stars, and then specifically address the topic 
of oscillation timing as a means to detect planets.
\par
In terms of successes to detect planets around evolved stars, the
subdwarf B stars as host stars stand out as a group. This class of
evolved objects will be introduced, and the difficulties in explaining
their sheer existence mentioned. Incidentally, the asteroseismology of
subdwarf B stars is also a very active field.
The \mbox{EXOTIME} planet searching program will be presented, which 
takes advantage of the long-term behaviour of these pulsations.
Current ideas on subdwarf B evolution, and the potentially crucial role
of planets, will be discussed, frequently resorting to the \object{V391
  Pegasi} system.
\begin{figure*}
\begin{center}
\includegraphics[width=\textwidth,angle=0]{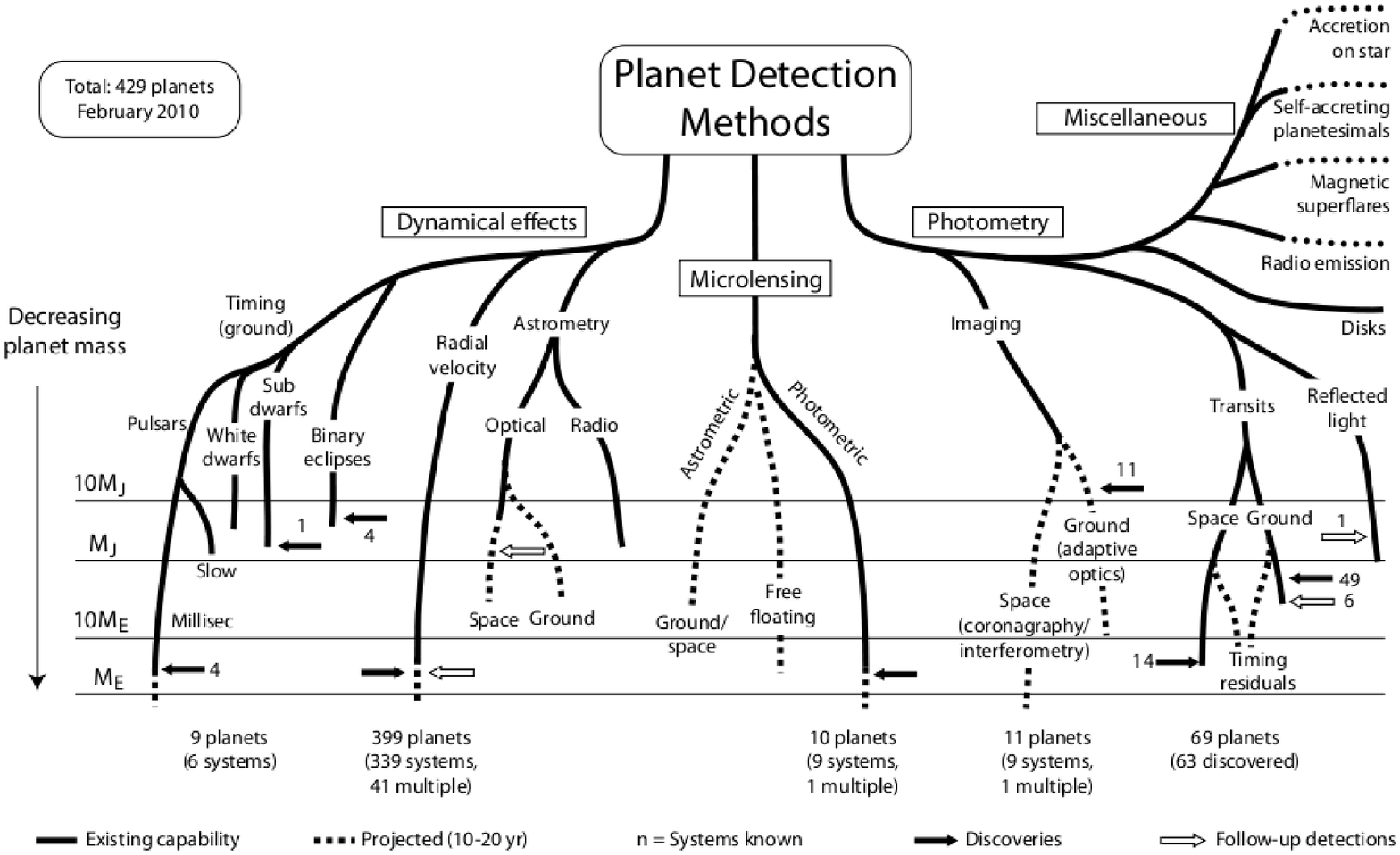}
\end{center}
\caption{The ''Perryman tree'': Detection methods for extra-solar
  planets with detectable masses on a (logarithmic) mass scale.
  Adopted from Figure~1 in \citet{2000RPPh...63.1209P}, updated to 
  include recent detections up to February 2010 
  (courtesy of M.\ A.\ C.\ Perryman). 
}
\label{fig:perrymantree}
\end{figure*}
\section{Extrasolar planet detection methods}
\label{sec:detectionmethods}
\subsection{Overview}
\label{subsec:overview}
The first extrasolar planet candidate was observed in
\citeyear{1989Natur.339...38L}, without however being claimed as such
at the time.\linebreak \citet{1989Natur.339...38L} instead suggested that the
sub-stellar companion they had found around \object{HD~114762}
probably was a brown dwarf: a class of intensely
searched-for objects, yet mostly elusive, at that period, and found to
be intrinsically rare in the role of companions to normal stars today.
\par
The first detection of an extrasolar planet around a solar-type star
properly published as a ''Jupiter-mass companion'' was by
\citet{1995Natur.378..355M}. This companion to \object{51 Pegasi} was
immediately confirmed by \citet{1995AAS...187.7004M}.
\linebreak 
\object{51 Pegasi b} constitutes the prototype of the hot Jupiters, but it
can also be more generally regarded as the prototype for all
extrasolar planets discovered with the radial velocity method.  While
the radial velocity method still is to be credited with the top score
in terms of number of planets (and multiple planetary systems)
discovered, other discovery methods gain their importance from the
fact that the observational biases involved can be significantly
different.
\par
$\!$%
Direct imaging, for instance, is obviously biased towards large
semi-major axes (e.g.\ \citealt{2008Sci...322.1345K}, 
\citealt{2008Sci...322.1348M}, and probably \citealt{2009A&A...493L..21L}), 
whereas radial velocity and transit measurements are
biased towards the detection of planets on orbits with small
semi-major axes (and large masses resp.\ radii). This correspondence
between the radial velocity and transit methods is in a sense a good
thing, since transit detections always need to be confirmed by radial
velocity measurements in order to secure a planet discovery, while
planet candidates from radial velocities detections usually remain
candidates as long as the inclination cannot be constrained. For
ground-based surveys, the micro-lensing technique is most sensitive in
the vicinity of the Einstein radius at 2\,-\,3~AU 
(\citealt{2009astro2010S..18B}; actual detections 
exist for semi-major axes in the range of 0.6\,-\,5.1~AU).
\par
The timing method, or more precisely, the various timing methods, also
constitute indirect methods.  While exploiting the same stellar
''wobbling'' effect induced by an unseen companion that is put to use
in the radial velocity method, it measures the varying light travel
time from the star(s) to the observer in the course of a ''wobbling''
cycle. Its amplitude (a direct measure of the projected semi-major
axis of the orbit followed by the central object) increases with large
companion masses, but also with large separations. Despite this
increasing sensitivity towards wider orbits, the fact that the
observational time base required for a detection increases for longer
orbital periods also must be factored into the overall detection
probability. Exactly as in the case of detections from radial velocity
variations, candidates discovered with any of the timing methods
suffer from a systematic uncertainty in the mass determination
whenever the orbital inclination remains unknown.
\par
A more detailed overview of the available planet detection methods
with an emphasis on their respective sensitivity to masses has been
given by \cite{2000RPPh...63.1209P}; an updated version with planet
detection counts up to early 2010 is given in
Fig.~\ref{fig:perrymantree}.
\subsection{Planetary systems around evolved stars}
\label{subsec:planetsaroundevolvedstars}
The majority of extrasolar planets known today were found around
solar-like stars. On the one hand, this has practical reasons - both
the number of lines in the optical and their sharpness decreases
towards earlier spectral types, while\linebreak later spectral types
tend to show augmented spectral variability due to activity, making
small periodic radial velocity signals harder to detect. On the other
hand, any dedicated searches for solar system analogues and Earth-like
planets will obviously primarily target solar-like stars. In the quest
to understand how our own solar system including the Earth has formed,
it seems plausible to assume that both the knowledge of the frequency
of similar systems as well as an overview of just how differently
planet formation has proceeded elsewhere are important ingredients.
\par
For stars evolving off the main sequence, the radial velocity method
remains applicable in the red giant regime. It has turned up a total
of 27 detections around G and K giants that add to the diversity of
systems known. Among the initial discoveries were those by
\citet{2002ApJ...576..478F},\linebreak
%
\citet{2005A&A...437..743H}, 
%
\citet{2006A&A...457..335H}, 
and\linebreak
\citet{2007A&A...472..649D}.
For all stars evolved beyond the first red giant branch, however, the
method that has most successfully been applied to detect extrasolar
planets so far is the timing method.
\subsection{Timing methods}
\label{subsec:timingmethods}
The timing method actually comes in a variety of flavours. In all
cases, a mechanism intrinsic to the central object provides a stable
clock the time signal of which reaches the observer with a delay or in
advance to the mean arrival times when a further body causes a cyclic
displacement (wobble) of the ''clock''. As a side note, the associated
change in period of the clock due to the Doppler effect is typically
much smaller than the light travel time delays.  
\par
The most prominent example for a central object's clock are the pulses
from rapidly rotating neutron stars. Pulsar planets were discovered in
this way to orbit \object{PSR 1257+12} by \citet{1992Natur.355..145W}, and
confirmed by \citet{1994Sci...264..538W}.
%
A further detection was reported for \object{PSR B1620-26} by
\citet{1993ASPC...36...11B}, \citet{1993ApJ...412L..33T}, and
\citet{1993Natur.365..817B}. 
This system in the globular cluster M4 is known for its assortment of
components, with a white dwarf in a tight orbit around the pulsar
(\citealt{1999ApJ...523..763T}; imaged using \textit{HST} by
\citealt{2003ApJ...597L..45R}), so the planet may be referred to as a
circumbinary planet.%
\footnote{%
As a side note, a different example for a hierarchical system is the
planet in the binary \object{$\gamma$ Cephei AB}, where the planet 
orbits the primary K subgiant component
(early speculations from radial velocity measurements by
\citealt{1988ApJ...331..902C} were first confirmed by
\citealt{2003ApJ...599.1383H}, and refined through a direct detection of
the secondary M dwarf component by \citealt{2007A&A...462..777N}).
}
\par
In a further approach, the timing method specifically targets
circumbinary planets by design. The clock in this case are frequent,
sharp eclipses in a close central binary system. Eclipse timing has
uncovered sub-stellar companions to subdwarf B stars (sdB, see
section~\ref{sec:hotsubdwarfs}) in HW~Vir-like binaries, in
(pre-)cataclysmic variables and possibly also in W~UMa systems.  Two
planetary companions have been accepted as confirmed around the
prototype system \object{HW~Vir} \citep{2009AJ....137.3181L}, while the
tertiary component to the binary \object{HS\,0705$+$6700}
\citep{2009ApJ...695L.163Q} probably lies in the mass range for brown
dwarfs.
Further HW~Vir systems with preliminary detections by
\citet{2010Ap&SS.tmp...50Q} include \object{HS\,2231$+$2441} and 
\object{NSVS~14256825}.
For the (pre-)cata\-clysmic variable central systems, tertiary
detections have been reported for the polar \object{DP~Leo}
\citep{2010ApJ...708L..66Q}, 
for the DA+dme binary \object{QS Vir} \citep{2010MNRAS.401L..34Q}, 
and perhaps for the nova \object{NN Ser} (\citealt{2009ApJ...706L..96Q}, 
status of the planet unconfirmed).%
\footnote{%
Provisional reports on more detections are also out on the W~UMa-type
eclipsing binaries \object{NY~Lyr} and \object{DD~Mon} 
\citep{2009PASA...26....7Q,2009PASJ...61..333Q}.}
\par
A variant of this method is to look for timing residuals in known
planetary transits in order to uncover additional planets in the
system.
\par
The third possibility is to resort to stellar oscillations as a
clock. This lead to the first discovery of a planet around a
subdwarf B star, \object{V391 Pegasi}, by \citet{2007Natur.449..189S}.  
The existence of this system suggested the possibility that a planet in an
orbit similar to that of the Earth may have survived the red giant
expansion of its presumably single host star, and has triggered
follow-up searches for comparable systems. I will come back to this
application in section~\ref{subsec:coherentoscillations}, and in more
detail in section~\ref{sec:exotime}.
\section{Stellar oscillation - extrasolar planet links}
\label{sec:stellaroscillationlink}
An obvious benefit of high-precision photometric time series of
planet-hosting stars (such as those obtained in ambitious transit
searches) is that the host star can additionally be characterized in
detail with asteroseismic methods if it oscillates.
\subsection{Asteroseismology}
\label{subsec:asteroseismology}
For small perturbations to a spherical equilibrium solution, an
infinite series of non-radial modes, with the radial modes included as
a special case, can occur (which can be described by nodal planes in
two angular directions, and spherical nodal
surfaces in the radial direction), leading to multi-periodic frequency
spectra. The equilibrium is restored by the actions of pressure and
buoyancy, with one of the two usually dominating in a particular
region of the star. The number of actually measurable frequencies
depends on\linebreak whether the corresponding modes are excited (i.e., if a
driving mechanism is converting radiative energy or, temporarily, convective
movement into pulsational kinetic energy), and on whether they are
observable as photometric or radial velocity variations (i.e., if the
mode geometry yields a detectable net effect in the integrated
observables).
\par
Given appropriate modelling capabilities, the density\linebreak structure of a
star can be inferred from observations of a sufficiently large number
of excited eigenmodes that probe the interior conditions
differentially. While the exact analysis approach can vary depending
on the class of variables considered, important fundamental parameters
that can be derived from this exercise are the stellar mass, radius
and (depth-dependent) chemical composition. 
\par
In solar-like pulsators, oscillatory eigenmodes are excited
stochastically by convection. The known pulsations are p modes
(acoustic modes, with displaced material restored by the action of
pressure), although g modes (gravity modes, action of buoyancy) are
also thought to exist in the deep interior of the sun. Considering
that derivatives of the local equilibrium gravity and radius can be
neglected in this case, approximations for high radial order can be
found. For acoustic modes described within the asymptotic theory, one
finds a regular frequency spacing for modes of the same low angular
degree corresponding to subsequent radial orders. The large frequency
separations, along with modifications introduced by considering the
effect of different degrees leading to additional small frequency
separations, together allow the definition of valuable diagnostic
tools.  This includes the famous {\'e}chelle diagrams, as well as the
'asteroseismic HR diagram' which allows to relate the two observables
large and small frequency separation directly to the masses and ages
of solar-like stars.
\subsection{Connection with solar-like oscillations}
\label{subsec:solarlikeoscillations}
The precise stellar parameters available through asteroseismic
investigations contribute to answering a number of key questions in
extrasolar planet research. 
One of the early\linebreak noteworthy examples was the idea to
investigate the scenarios for the origin of the enhanced metal content
of \object{$\mu$~Arae} (\object{HD~160691}), host to a system of at least four planets,
with asteroseismic methods. From ground-based radial velocity observations
\citep{2005A&A...440..609B}, \citet{2005A&A...440..615B} attempted to
decide whether the overmetallicity was limited to the outer layers
(accretion scenario) or was present\linebreak throughout the star (resulting
from enhanced metallicity in the original proto-stellar cloud). The
latter scenario with enhanced planet formation rates in intrinsically
metal-richer star-forming clouds has now been generally accepted, a
result incorporated into the more recent analysis by\linebreak
\citet{2009arXiv0903.5475S} that has in addition uncovered a high helium
abundance. It will remain to be seen if contributions from
asteroseismology will also be able to help solve the possibly related
mystery of enhanced lithium depletion in planet-hosting stars
\citep{2009Natur.462..189I}.
Fundamental links between the two programmes of the space mission
\textit{Corot}, the exoplanet search and the asteroseismology programme, have
been pointed out by\linebreak \citet{2006ESASP1306...77V} (see also
\citealt{2007A&A...471..885S,2008arXiv0809.0249V}).
In \textit{Corot} these two programs are conducted with separate pairs of
detectors operated with different instrumental setups: an on-focus
setup with dispersion through a bi-prism sampled every 512\,s in the
exoplanet field, and a highly out-of-focus setup sampled every 1\,s in
the seismology field.
\par
While a smaller technical dichotomy continues to exist with the long
(30\,min) and short (1\,min) cadence readout modes for the individual apertures on the
\textit{Kepler} satellite's detectors, the boundaries start to dissolve here
since the sampling is at best loosely associated with the classification
of a target as belonging to the planet-hunting core program or a
program such as the asteroseismic investigation, and can simply be
re-assigned. \textit{Kepler} has demonstrated its capabilities early on in the
mission schedule with observations of the known transiting exoplanet
host \object{HAT-P-7} \citep{2009Sci...325..709B}. \object{HAT-P-7} has in the meantime
been further characterized through an analysis of its simultaneously
discovered solar-like oscillations\linebreak \citep{2010arXiv1001.0032C}.
\par
\textit{Kepler} has now also delivered its first five genuine extrasolar planet
discoveries \citep{2010Sci...327..977B}. The routine analysis of
solar-like oscillations in newly-discovered planet-hosting stars, as
well as the analysis of a large variety of asteroseismology targets of
interest to the pulsation science community, have been
institutionalized in the \textit{Kepler Asteroseismic Investigation} (KAI) and
the \textit{Kepler Asteroseismic Science Consortium}
(KASC, see\linebreak \citealt{2008JPhCS.118a2039C}, and in particular the first
results in \citealt{2010PASP..122..131G} and\linebreak
\citealt{2010arXiv1001.0032C}). The importance of determining the
stellar radius as accurately as possible stems from the circumstance that a
transit light curve yields the radius \emph{ratio} between star and
planet. In order to derive the transiting planet's absolute radius the
value of the stellar radius must be known. Together with the planet
mass determined from the confirmation radial velocity curve (again,
the stellar mass must be known), the planet's mean density is found.
In the proposed \textit{Plato} mission\linebreak
\citep[e.g.][]{2009platoassessmentstudy},\linebreak the high-precision
determination of planet host star radii and other
fundamental stellar parameters from asteroseismology is an integral
part of the planet hunting and characterization concept.
\subsection{Connection with coherent oscillations}
\label{subsec:coherentoscillations}
The above considerations could in principle with the same rationale be
extended to planet-hosting stars that exhibit\linebreak pulsations
other than solar-like. Well beyond this, the striking capabilities in
particular of \textit{Kepler} evidently open many possibilities for
genuine asteroseismological applications\linebreak that target questions in many
areas of stellar astrophysics. This includes the classical pulsators
which, instead of being stochastically excited as the
solar-like stars, exhibit unstable modes driven by the $\kappa$
mechanism, with topical applications extensively described by
\citet{2010PASP..122..131G}.
\par
The standard asteroseismic exercise derives the instantaneous
structure of a star, often relying on model structures from full
evolutionary calculations in the process. As an extension, period
changes in the oscillatory eigenfrequencies due to evolutionary
effects can be considered as an additional constraint to find the best
solution in parameter space. Given a series of evolutionary models
already subjected to a stability analysis, the rate of change
$\dot{P}$ of modes in a specific model can be determined without too
much trouble. Due to the typically very long evolutionary time scales
involved, measuring the secular evolution in real stars is
observationally expensive and somewhat more complex.
\par
An example of a class of objects where pulsators can be found that
prove to be coherent and stable on time scales of many years are the
ZZ~Ceti variables on the DA white dwarf cooling track. The pulsations
in these objects are due to the recombination of hydrogen in a narrow
temperature range, leading, via the associated increase of the opacity
in the outer layers, to the manifestation of low-degree low-order g
modes. In contrast to the millisecond pulses in spun-up pulsars, and
also in comparison to the pulse duration in normal ''slow'' pulsars,
the pulsation periods in white dwarfs are much longer: of the order of
a few minutes. The precision that can be reached in determining the
clock rate is hence accordingly lower, and long-term changes are more
readily analyzed using O$-$C techniques, instead of determining
$\dot{P}$ directly as the derivative of a series of
quasi-instantaneous $P$ measurements over time.
\par
Measurements of $\dot{P}$ and its interpretation in the context of
cooling times exist for a small number of suitable pulsating white
dwarfs: \object{PG\,1159$-$035}
(\citealt{1999ApJ...522..973C};
\citealt{2008A&A...489.1225C}),
\object{G117-B15A}
(\citealt{1991ApJ...378L..45K};\linebreak
\citealt{2000ApJ...534L.185K};
\citealt{2005ApJ...634.1311K}),
and a larger sample of a total of 15 objects investigated by
\citet{2008ApJ...676..573M}.
It was quickly recognized \citep[e.g.\ by][]{1997ASPC..119..123P} that
the influence of a possible unseen companion can be measured as a side
effect from the same data. Applying the ideas of the timing method
(section~\ref{subsec:timingmethods}) has given the long-term
photometric monitoring of pulsations in white dwarfs and related
objects a new spin as a means to search for planets around evolved
stars. With respect to previous work, the efforts by
\citet{2008ApJ...676..573M} show a shift of focus to that effect.
Their proposed planet candidate around \object{GD 66} has however
remained unconfirmed.
\par
The same measurements are possible for a different\linebreak group of
compact oscillators, the subdwarf B stars (see below), a
quantitatively uncommon feeder channel for white dwarfs.  The rapid
pulsations in subdwarf B stars, of the order of minutes just as in the
ZZ~Ceti white dwarfs, are generated via a $\kappa$ mechanism,
providing potentially suitable conditions for reasonable long-term
coherence. Yet, results have only been published for one pulsating
subdwarf B star so far, \object{V391 Pegasi} (aka
\object{HS\,2201$+$2610}).  As stated in
section~\ref{subsec:timingmethods}, these measurements, simultaneously
to $\dot{P}$, revealed the presence of a companion with a planetary
mass. The possible reasons for this initially unexpected, instant
success, sometimes jokingly referred to as ''100\% discovery rate'',
are worth a more in-depth investigation.
\section{Hot subdwarf stars}
\label{sec:hotsubdwarfs}
\subsection{Evolution}
\label{subsec:evolution}
Subdwarf B stars (sdBs) are subluminous hot stars that are found in an
effective temperature range from $20\,000\,\textrm{K}$ to
$40\,000\,\textrm{K}$ at surface gravities between about $5.0$ and
$6.2$ in $\log{(g/\textrm{cm\,s}^{-2})}$, and that can in many cases
be identified with evolved stellar models on the extreme
horizontal\linebreak branch (EHB). Their masses are expected to peak
around the value for the \element{He} core flash at
$0.46\,\textrm{M}_{\odot}$. As is true for all horizontal branch stars, extreme
horizontal branch stars have a \element{He} burning core but, due to
previous significant mass loss, no \element{H}-shell burning in their
thin hydrogen shells. The thinness of the shell leads to their blue
appearance, so that the flux from hot subdwarfs (including the hotter
sdOs) contributes significantly to the UV excess observed in galaxy
bulges and elliptical galaxies. A great recent review of the
observational properties of hot subdwarfs including
binarity, kinematics, as well as
current modelling capabilities to describe their
atmospheres, interiors, and evolutionary history, can be found in
\citet{2009ARA&A..47..211H}.
The most puzzling question about hot subdwarfs remains what the precise
evolutionary status of these core-helium burning or even more evolved
objects really is. Here I only focus on how to possibly produce the sdB type.
The basic problem in the standard single-star evolutionary scenario is to shed
the hydrogen envelope almost entirely just before or at the moment of the
\element{He} flash. This would require \textit{ad hoc} strong mass loss
through stellar winds for a certain fraction of stars upon reaching the tip of
the first giant branch. Explaining the formation of subdwarf B stars in the
context of binary evolution is therefore now generally favoured over
single-star scenarios \citep{2002MNRAS.336..449H,2003MNRAS.341..669H}.
\par
Following the overview by \citet{2008ASPC..392...15P}, three genuine
binary formation scenarios, and additionally the merger scenario, can be
distinguished. All three of the binary scenarios involve Roche-Lobe
overflow\linebreak (RLOF), each at different stages and under different
conditions.
\par
In the ''stable RLOF + CE'' channel, the system initially goes through a first
mass-transfer phase with stable RLOF that turns the evolving component into a
\element{He} white dwarf. When the second component, the future subdwarf B
star, then evolves to become a red giant, the second mass-transfer phase
can happen dynamically. This unstable RLOF, where the matter transferred cannot
all be accreted by the \element{He} white dwarf, leads to the formation of a
common envelope (CE). After spiral-in and envelope ejection, the resulting
system consists of the \element{He} white dwarf and the sdB -- the core of the
giant with its envelope removed -- in a short-period binary ($<$ 10 days).
\par
In the ''CE only'' channel, unstable RLOF occurs when the future subdwarf B
star starts transferring matter to a lower-mass main-sequence companion near the
tip of the first giant branch. The ensuing common-envelope phase\linebreak again leads 
to a closer final configuration of the resulting system, which will consist of a
low-mass main sequence star and the sdB in a short-period binary ($<$ 10 days).
\par
Besides these two common-envelope channels (involving unstable RLOF), the
''stable RLOF'' channel can also produce sdBs. Stable Roche-Lobe overflow can
occur when the future subdwarf B star starts transferring matter to a main
sequence companion at mass ratio below $\sim$ 1.2. As before, this happens
near the tip of the red giant branch, but this time the system more likely
widens due to the mass transfer. The resulting system consists of a
main sequence or subgiant star and the sdB in a wide binary
(${\mathrel{\hbox{\rlap{\hbox{\lower4pt\hbox{$\sim$}}}\hbox{$>$}}}}$ 10 days).
\par
The problem with all of the above binary scenarios is that (apparently?)
single sdBs also exist. In addition to the single star scenario involving
variable mass loss as mentioned above, a further possibility to produce these
are\linebreak mergers.
In the ''merger'' scenario, two \element{He} white dwarfs in a close
system, produced while undergoing one or two common envelope phases,
spiral towards each other due to angular momentum loss via
gravitational radiation until the less massive one gets disrupted and
its matter accreted onto the more massive component. At a critical
mass, the accretor can ignite helium fusion and the merger product
would hence indeed turn into a single sdB.  
\par
In the confrontation with observations, both the types of companion
found in close systems (white dwarf or low-mass main-sequence stars)
as well as the close binary frequency are roughly as expected when the
above merger\linebreak channel is considered. Yet observationally, the
mass spectrum for the companions is broader than expected from the
standard formation scenarios for sdB+WD, sdB+dM and single sdBs. On
the low-mass end, it can be argued to comprise the \object{V391
  Pegasi} planet and, possibly, further\linebreak secondary
sub-stellar companions. Four companions with unusually high masses
have also been reported; according to
\citet{2008MmSAI..79..608G,2009JPhCS.172a2008G,2010Ap&SS.tmp...84G},
the massive compact companions found from radial velocity variations
must at least be heavy white dwarfs or in two cases even neutron stars
or black holes.
\par
While the problem of sdB formation has not been fully solved, the
subsequent evolution of a subdwarf B star towards the white dwarf
cooling sequence is more straightforward. One interesting
characteristic is that this evolution, due to the thinness of the
outer layers, bypasses the asymptotic giant branch, as was noted early
on by \citet{dorman:93}. Another aspect is that the sdB+dM close
systems constitute potential progenitors for cataclysmic variables.
Furthermore, given the variety of configurations in which sdBs are
found as one component, the diversity of possible progeny systems is
not restricted to pre-CVs.
\par
All of the above, while far from understood in the details, implies
that the sdB's stellar core as it looks like after the first giant branch
-- or after a merger -- is \emph{almost} laid bare. It is hence very well
accessible to asteroseismological methods.
\subsection{Pulsating subdwarf B stars}
\label{subsec:pulsatingsdBs}
Only a small fraction of the sdBs show pulsational variations, with
non-pulsators also populating the region in the HRD where the
pulsators are found. There are p- \mbox{(pressure-)} mode and 
g- (gravity-)mode types of pulsation. 
\par
The rapidly pulsating subdwarf B stars (sdBV$_{\textrm{r}}$) were
discovered observationally by \citet{1997MNRAS.285..640K}. The short
periods of these p-mode pulsators are of the order of minutes and have
amplitudes of a few tens \textrm{mmag}. These pulsations were independently
predicted to exist by\linebreak \citet{1997ApJ...483L.123C}, who in this initial
and subsequent research papers explain the driving to result from a
$\kappa$ mechanism due to a Z-opacity bump accumulated by radiative
levitation.
\par
The longer periods in the slowly pulsating subdwarf B stars
(sdBV$_{\textrm{s}}$) range from $30$ to $80\,\textrm{mins}$ at even
lower amplitudes of a few \textrm{mmag}. This group of g-mode pulsators was
discovered by \citet{2003ApJ...583L..31G} and has been explained to
pulsate due to the same $\kappa$ mechanism as the sdBV$_{\textrm{r}}$
type by \citet{2003ApJ...597..518F}.
\par
A number of objects are known that show both types of mode
simultaneously and are referred to as hybrid pulsators, or
sdBV$_{\textrm{rs}}$ in the nomenclature of
\citet{2010IBVS.5927....1K} that is used here. These hybrids lie in
the overlapping region between the hotter short-period and the cooler
long-period pulsators. The prototype and a prominent example is
\object{HS\,0702$+$6043}. Its rapid pulsations were discovered by
\citet{2002A&A...386..249D}, and subsequently the slower ones by
\citet{2005ASPC..334..530S,2006A&A...445L..31S}. Further examples can
be found in \citet{2008JPhCS.118a2015S} and references therein.  It
should however be noted here that the planet host star \object{V391
  Pegasi}\linebreak (\object{HS\,2201$+$2610}) is also among the hybrids. It has
five p modes (\citealt{2001A&A...368..175O};
\citealt{2002A&A...389..180S}) and at least one g mode
\citep{2008ASPC..392..339L,2009A&A...496..469L}.
\par
Exploiting the sdBV$_{\textrm{r}}$ asteroseismologically has the
potential to test diffusion processes and has lead to a first mass
distribution for subluminous B stars
(\citealt{2008CoAst.157..168C,2008ASPC..392..231F,2009CoAst.159...75O} and references
therein.) On the other hand, the extent of the instability region for
the sdBV$_{\textrm{s}}$, but in particular the existence of the
sdBV$_{\textrm{rs}}$, have challenged details of the input physics for
the models. The actual composition of the 'Z' in the Z bump (iron,
nickel) as well as the role of opacities have been discussed in this
context \citep{2006MNRAS.372L..48J,2007MNRAS.378..379J}. While, quite
similarly to the findings for the sdB variables, using updated Opacity
Project (OP) instead of OPAL opacities also improves the situation in
$\beta$ Cep and [SPB] variables, the opposite is the case in the sun
with helioseismology. The most appropriate opacities are therefore
still under discussion.
\par
Further open questions in this field are the co-existence of pulsators
and non-pulsators, and also the origin of the amplitude variability
observed in a number of sdBVs\linebreak \citep{2010Ap&SS.tmp...82K}.
\subsection{Sub-stellar companions of subdwarf B stars}
\label{subsec:substellarcompanions}
When a companion with planetary mass was found around the hybrid
pulsating subdwarf B star \object{V391 Pegasi} with the timing method,
this indicated that a previously undiscovered population of
sub-stellar companions to apparently single subdwarf B stars might
exist \citep{2002A&A...389..180S,2007Natur.449..189S}.

\par
As for low-mass tertiary bodies around HW~Vir-like systems detected by
eclipse timing, \citet{2009AJ....137.3181L} have reported two planetary
companions around the system\linebreak \object{HW~Vir}
itself. \citet{2009ApJ...695L.163Q} put the mass of the tertiary body
in the \object{HS\,0705$+$6700} system in the range for brown
dwarfs. Still unconfirmed detections by \citet{2010Ap&SS.tmp...50Q}
exist for \object{HS\,2231$+$2441} and \object{NSVS~14256825}.
\par
A different type of detection has been put forward by
\citet{2009ApJ...702L..96G,2010Ap&SS.tmp...84G}, who report
measurements of radial velocity variations in \object{HD 149382}
indicative of a planetary companion, variations which if real are
indicative of a close sub-stellar companion in a 2.4~d orbit.  Based
on the above and this finding (that yet has to be confirmed), these
authors also argue for a decisive influence of sub-stellar companions
on the late stages of stellar evolution. The \mbox{EXOTIME} program
aims to increase the empirical data available on which to base such
discussions.
\par
\begin{figure*}
\begin{center}
\includegraphics[height=0.8\textwidth,angle=90]{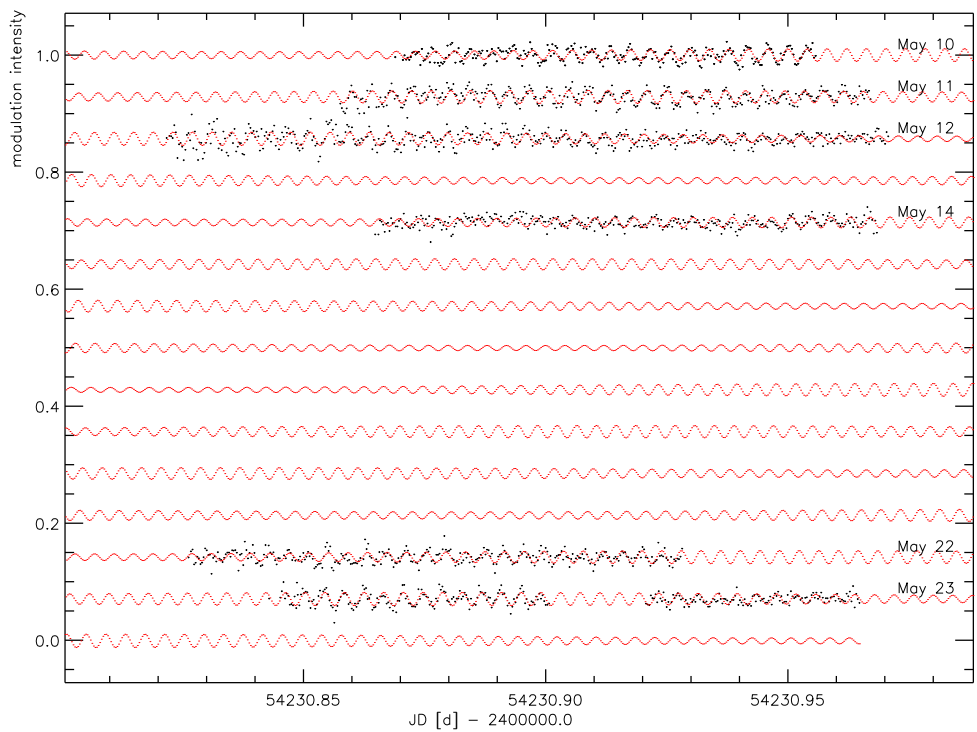}
\end{center}
\caption{Photometric data obtained with the $1.2\,\textrm{m}$
  MONET/North telescope during a period of $2~\textrm{weeks}$ in May
  2005, showing the pulsations in the light curve of \object{V391
    Pegasi}. Starting from a mean intensity of unity on the first
  night, subsequent observations are shifted downward by a fixed
  offset each night. The actual data points (black dots) are overlaid
  with model for the pulsations in red. One such run spanning several
  nights is required to derive one O$-$C point by comparing the
  current phasing to that of a mean model.%
}
\label{fig:hs2201monet}
\end{figure*}
\section{The EXOTIME program}
\label{sec:exotime}
\subsection{The planet-hosting pulsating sdB V391 Pegasi}
\label{subsec:planethostingv391pegasi}
\object{V391 Pegasi} (\object{HS\,2201$+$2610}) was first discovered
to be a rapidly pulsating subdwarf B star by
\citet{2001A&A...368..175O}.$\!$ Additional slow pulsations were
subsequently reported by \citet{2009A&A...496..469L}.
\citet{2002A&A...389..180S,2007Natur.449..189S} were able to derive
$\dot{P}$ values for the two strongest pulsation modes, and found an
additional pattern in the observed--calculated (O$-$C) diagrams that
revealed the presence of a giant planet in a $3.2\,\textrm{year}$
orbit. The fact that this cyclic variation has been measured
independently from two frequencies considerably strengthens the
credibility of this discovery. Actually, \citet{2007Natur.449..189S}
detected parabolic and sinusoidal variations in the O$-$C diagram
constructed for the two main pulsation frequencies at
$349.5\,\textrm{s}$ and $354.2\,\textrm{s}$ over the observing period
of seven years. The sinusoidal component with its $3.2\,\textrm{year}$
periodicity is attributed to the presence of the very low-mass
companion \object{V391 Pegasi b} at $m\sin{i}=3.2 \pm
0.7\,\textrm{M}_{\textrm{Jup}}$. The scenarios proposed for the origin of this
planet are discussed in section~\ref{subsec:fateofplanets}.
\subsection{Further characterization of V391 Pegasi system}
\label{subsec:characterizingv391pegasi}
The determination of the true mass of the ''asteroseismic planet''
\object{V391 Pegasi b} requires to find a constraint on the orbital
inclination. Besides the orbital inclination, the orbit eccentricity
has also not been well determined so far. The constraints on possible
further planets are weak and currently allow for a second planet in
the system massive enough to be detected. Continued photometric
monitoring (on-going, see for example Fig.~\ref{fig:hs2201monet},
Silvotti et al.\ \textit{in prep.}) will be able to:
check if the O$-$C evolves as predicted from the orbital solution,
investigate the eccentricity, investigate a possible multiplicity,
attempt another independent re-detection of the planet from $\dot{P}_3$,
and search for rotational splitting in the pulsations (see below).
\par
As a first step to derive the mass of the known companion object with
follow-up observations, \citet{2009CoAst.159...91S} have attempted to
determine the orbital inclination from\linebreak spectroscopy.  The
approach suggested to achieve this is to use the stellar inclination
as a primer for the orbital orientation. ''Stellar inclination'' can
refer to the rotational or the pulsational axis, which as a further
necessary simplification are assumed to be aligned, and can in turn
then be derived by combining measurements of $v_\textrm{rot}$ and
$v_\textrm{rot}\sin{i}$.
\par
\begin{figure*}
\begin{center}
\includegraphics[width=0.8\textwidth,angle=0]{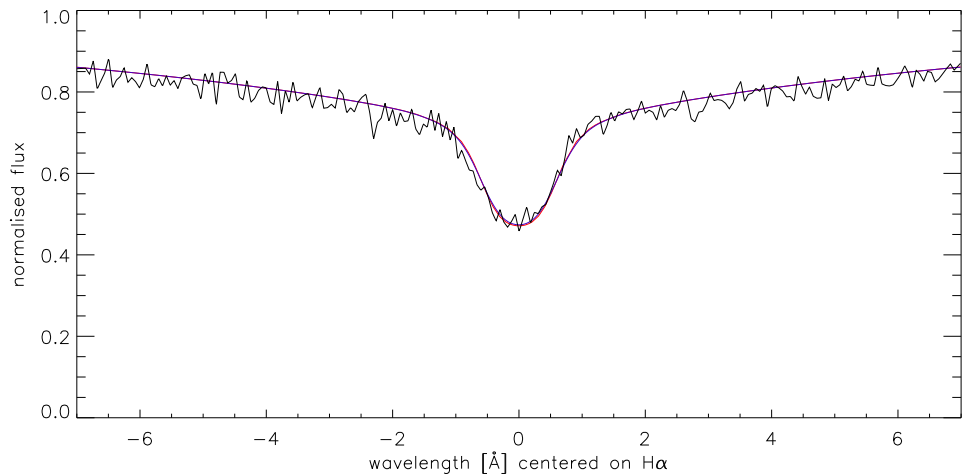}
\end{center}
\caption{Detail of the observed \element{H}$\alpha$ core in a
  pulsation-integrated Keck spectrum of \object{V391 Pegasi}. The data
  (shown as a black line) are plotted centered on 6562.798\AA. While
  measurements on the \element{H}$\alpha$ core may be subject to
  systematic effects, this line is the only one available for such
  investigations in the spectral range covered by the observation.
  The flux was normalized; it does not reach unity at the 
  edges of the plot as these regions are still well within the line wings. 
  Overlaid are 1) 
  a model spectrum in red calculated from a NLTE pure H atmosphere model at 
  $T_{\textrm{eff}}=30\,000\,\textrm{K}$ and 
  $\log{(g/\textrm{cm\,s}^{-2})}=5.50$,
  and 2) 
  a model spectrum in blue from the same atmosphere model as before but
  rotationally broadened with $0.2\,\textrm{\AA}$ corresponding to
  $9\,\textrm{km\,s}^{-1}$.
}
\label{fig:keck_halphacore}
\end{figure*}
\par
The value for $v_\textrm{rot}$ is in principle accessible through
rotational splitting in the photometric frequency spectrum\linebreak
(which has however not been found for \object{V391 Pegasi} yet), while
the projected rotational velocity $v_\textrm{rot}\sin{i}$ can be
measured from the rotational broadening of spectral lines. This
rotational broadening must be deconvolved from the additional
pulsational broadening caused by the surface radial velocity variation
in high S/N phase averaged spectra.
\par
Both phase averaged and phase resolved high resolution {\'e}chelle spectra
were obtained in May and September 2007 with the Hobby-Eberly
Telescope (HET), and one phase averaged spectrum in May 2008 with the
Keck~1 telescope, in order to put limits on the pulsational radial
velocities. {\'E}chelle spectra of \object{V391 Pegasi} were taken during
May and September 2007 with the HRS ($R=15\,000$) of the HET at the
McDonald Observatory, and with HIRES ($R=31\,000$) at the Keck~1
telescope atop Mauna Kea in May 2008.
\par
Using standard data reduction procedures, the individual {\'e}chelle
orders were merged and the final spectra carefully normalized and
finally summed. This results in a set of individual spectra
($S/N \approx 3$), in particular two September 2007 high time
resolution series, and summed spectra for May and September 2007
\citep{2008CoAst.157..325K}.
\par
In an attempt to ''clean'' the relevant rotational broadening from
pulsational effects, the spectra in September obtained in time
resolved mode were combined to a series of ten phase resolved
averaged spectra ($S/N \approx 9$) for the main pulsation period of
$349.5\,\textrm{s}$ (similar to \citealt{2007A&A...473..219T}).
\par
The cross-correlation of this series of averaged spectra with a pure
hydrogen NLTE model spectrum at $T_{\textrm{eff}}=30\,000\,\textrm{K}$
and $\log{(g/\textrm{cm\,s}^{-2})}=5.5$ as a template yields
pulsational radial velocity measurements for the different Balmer lines.
The maximum amplitude of a sinusoidal curve (fixed at the expected
period) that could be accommodated in comparison to the
weighted means of the Balmer lines reveals that any pulsational radial velocity amplitude
is smal\-ler than the accuracy of our measurements and confirms the
upper limit of $16\,\textrm{km\,s}^{-1}$ given by
\citet{2008CoAst.157..325K}.
\par
The resolution of the model template matches the spectral resolution
of the (pulsation-averaged) Keck spectrum. A comparison of the
\element{H}$\alpha$ NLTE line core shape (see
Fig.~\ref{fig:keck_halphacore}) yields a more stringent upper
limit for the \emph{combined} broadening effect of pulsation and
rotation of at most $9$\,{km\,s$^{-1}$},
meaning better spectral resolution and signal
to noise data will be necessary to measure $v_\textrm{puls}$ and
$v_\textrm{rot}\sin{i}$.
\subsection{Exoplanet Search with the Timing Method}
\label{subsec:exotime}
\begin{table*}[t]
\begin{center}
\small
\caption{Overview of the EXOTIME targets, see section~\ref{sec:exotime} for details.}
\label{tbl:targets}
    \begin{tabular}{lllll}
      \hline
      target             & coordinates (equinox 2000.) & $m_{\textrm{B}}$ & status & \\
      \hline
      \object{HS\,2201$+$2610}    & 22:04:12.0 \quad $+$26:25:07 &14.3& {collecting data} &
                                     {aka \object{V391 Pegasi}},
                                     {planet candidate},
                                     {$\sin{i}$~unknown}  \\
      \object{HS\,0702$+$6043}    & 07:07:09.8 \quad $+$60:38:50 &14.7& {collecting data}&{see \citet{2010Ap&SS.tmp....1S}}\\
      \object{HS\,0444$+$0458}    & 04:47:18.6 \quad $+$05:03:35 &15.2 & {collecting data}&{see \citet{2010Ap&SS.tmp....1S}}\\
      \object{EC\,09582$-$1137}~& 10:00:41.8 \quad $-$11:51:35 &15.0& {collecting data}& \\
      \object{PG\,1325$+$101}     & 13:27:48.6 \quad $+$09:54:52 &13.8& {collecting data}& \\
      \hline
    \end{tabular}
\end{center}
\end{table*}
Following the serendipitous discovery of \object{V391 Pegasi b}, the
\mbox{EXOTIME}\footnote{\texttt{http://www.na.astro.it/$\sim$silvotti/exotime/}}
monitoring program was set up to search for similar systems.
\mbox{EXOTIME} monitors the pulsations of a
number of selected rapidly pulsating subdwarf B stars on time-scales
of several years with the immediate observational goals of
a) determining $\dot{P}$ of the pulsational periods $P$ and
b) searching for signatures of sub-stellar companions in O$-$C residuals due
to periodic light travel time variations, which would be tracking the central star's
companion-induced wobble around the centre of mass.
\par
The long-term data sets should therefore on the one hand be useful to
provide extra constraints for classical asteroseismological exercises
from the $\dot{P}$ (comparison with ''local'' evolutionary models),
and on the other hand allow to investigate the preceding evolution of
apparently single sdB targets in terms of possible ''binary''
evolution by extending the otherwise unsuccessful search for
companions to potentially very low masses.
\par
As noted before, timing pulsations to search for companions samples a
different range of orbital parameters, inaccessible through orbital
photometric effects or the radial velocity method: the latter favours
massive close-in companions, whereas the timing method becomes
increasingly more sensitive towards wider separations. A further
advantage of timing versus radial velocities is that the former,
although observationally expensive, is easier to measure than the
latter. In fact, it is very hard to achieve the required
accuracy in radial velocity measurements from the few and broad lines
in hot subdwarf stars
(a notable exception is the publication of small RV variations in the
bright sdB HD~149382 as reported by \citealt{2009ApJ...702L..96G}.)
\par
The targets selected for monitoring in the \mbox{EXOTIME} program are
listed in Table~\ref{tbl:targets}. The target selection criteria
applied to compile this list over time have been described by
\citet{2010Ap&SS.tmp....1S}.
\par
\object{V391 Pegasi} (\object{HS\,2201$+$2610}) appears as the first
entry, as the monitoring of this system is on-going (see
\ref{subsec:planethostingv391pegasi}). The rapid pulsations in
\object{HS\,2201$+$2610} were discovered by
\citet{2001A&A...368..175O}, and additional slow pulsations by
\citet{2009A&A...496..469L}. An asteroseismology analysis of the star
is included in \citet{2002A&A...389..180S,2007Natur.449..189S}.
\par
Rapid oscillations were discovered in the second target on the list,
\object{HS\,0702$+$6043}, by \citet{2002A&A...386..249D}, and
simultaneous slow oscillations were reported by
\citet{2006A&A...445L..31S}. The on-going \mbox{EXOTIME} observations
for\linebreak \object{HS\,0702$+$6043} have also previously been summarized by
\citet{2008CoAst.157..185L,2009CoAst.159...94L} and include a
significant contribution of data by \citet{2010ApSS.F}.
\par
The target \object{HS\,0444$+$0458} was first discovered to pulsate by
\citet{2001A&A...378..466O}, and has been further characterized by
\citet{2007MNRAS.378.1049R}. \mbox{EXOTIME} has followed it regularly
since 2008.
\par
\citet{2006MNRAS.367.1603K} discovered rapid pulsations in\linebreak
\object{EC\,09582$-$1137} which \citet{2009A&A...507..911R} subjected
to an asteroseismology analysis. \object{EC\,09582$-$1137} is included
in \mbox{EXOTIME} as a southern hemisphere target.
\par
The discovery of pulsations in \object{PG\,1325$+$101} by\linebreak
\citet{2002A&A...383..239S} was also taken advantage of by
characterizing the star asteroseismologically\linebreak
\citep{2006A&A...459..557S,2006A&A...459..565C}. This target is also
extensively being observed within \mbox{EXOTIME}.
\par
\citet{2010Ap&SS.tmp....1S} present a portion of the observations
currently available, describe the treatment of the data and display
the first, still relatively short, O$-$C diagrams for the \mbox{EXOTIME}
targets \object{HS\,0444$+$0458} and \object{HS\,0702$+$6034}.\linebreak 
Not surprisingly, these illustrate the need for further observations
and interpretation. The analysis resorts to tools provided by
\citet{1999DSSN...13...28M} and\linebreak \citet{2005CoAst.146...53L}.
\begin{figure*}
\begin{center}
\includegraphics[width=0.8\textwidth,angle=0]{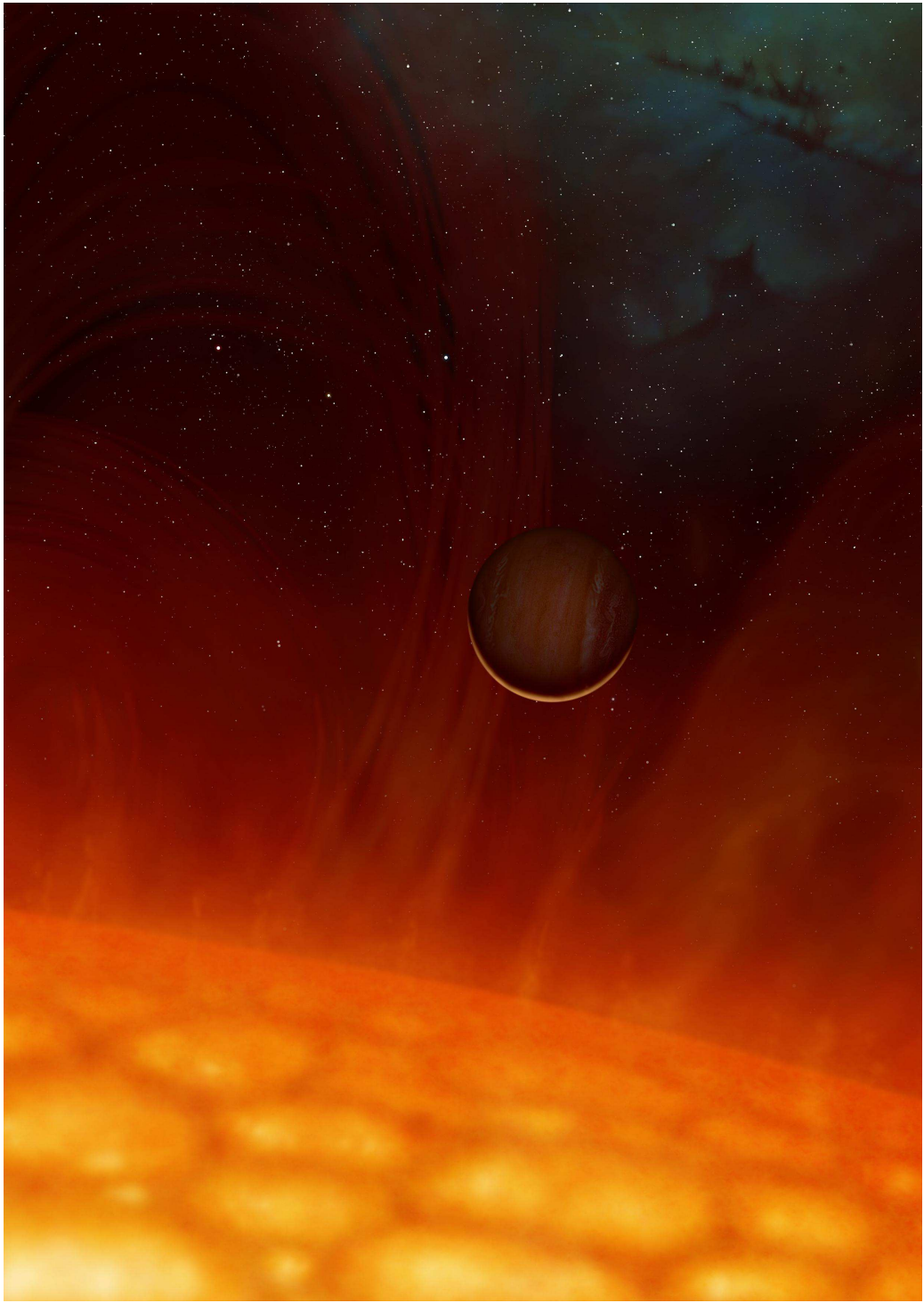}
\end{center}
\caption{The subdwarf~B (sdB) star \object{V391 Pegasi} oscillates in
  short-period p modes and long-period g modes, making it one of the
  known hybrid pulsators among sdBs. As a by-product of the effort to
  measure secular period changes in the p modes due to evolutionary
  effects on a time scale of almost a decade, the O$-$C diagram has
  revealed an additional sinusoidal component attributed to a periodic
  shift in the light travel time caused by a planetary-mass companion
  around the sdB star in a $3.2\,\textrm{year}$ orbit. In the above
  artistic impression, the \object{V391 Pegasi} system is shown at an
  earlier evolutionary stage in one of the proposed scenarios where,
  roughly $10^8\,\textrm{years}$ ago, the star, at maximum red giant
  expansion, almost engulfed the planet.
  \textcopyright\
  Image courtesy of HELAS, the European Helio- and
  Asteroseismology Network, funded by the European
  Union under Framework Programme 6; Mark Garlick, artist.
}
\label{fig:hs2201}
\end{figure*}
\section{Sub-stellar companions of evolved stars}
\label{sec:substellarcompanions}
\subsection{Post-main-sequence evolution}
\label{subsec:postmainsequenceevolution}
For single stars with initial masses below $\approx
2.2\,\textrm{M}_{\odot}$ on the main sequence, the canonical steps in
the evolution are: hydrogen shell burning following core hydrogen
exhaustion, core mass increase (due to the shell burning) and core
contraction with increasing electron degeneracy simultaneous to radius
swelling on the first giant branch (RGB), ended by onset of a core
helium flash, which brings the star onto the horizontal branch. Their
masses now lie in the range roughly
$0.5\,\textrm{M}_{\odot}<M<1.0\,\textrm{M}_{\odot}$, with a canonical
core mass of $0.46\,\textrm{M}_{\odot}$ and variable envelope masses
with or\linebreak without hydrogen shell burning that determine the location on
the horizontal branch. The hottest, bluest objects are the
least massive and have thin envelopes that put the most extreme of
these objects with inert hydrogen shells on the EHB. This sequence is
terminated at the point where the helium main sequence coincides with
the horizontal branch helium core mass.
\par
After core helium exhaustion the complex asymptotic giant branch
evolution with \element{H} and \element{He} shell burning, thermal
pulses, convection and dredge-up processes, and extreme mass loss
unfolds. The subsequent post-AGB phase sees the creation of a
planetary nebula from the envelope material lost and the emergence of
the stellar remnant on its way to the white dwarf cooling track.
\par
As the subdwarf B stars are in between the RGB and AGB giant expansion
phases, or more precisely, will not undergo AGB evolution, their
observation allows to separate the effects of the first giant branch
on planets from those of the asymptotic giant branch.
\subsection{The fate of planets around evolving stars}
\label{subsec:fateofplanets}
The starting point of discussions on planets around evolved stars is
often that relic planets around white dwarfs are expected simply due to
the considerable number of planets found around main sequence stars.
The question then is at what initial masses and orbital separations
the planets actually can survive both giant expansion phases of their
host star without being engulfed, disrupted, evaporated or\linebreak ejected.
\citet{2002ApJ...572..556D} consider the effects of mass loss,
planet-planet interactions, and orbit stability and conclude
that while inner planets will perish, far-out small bodies
and distant planets have the potential to create a new dust disk. Such
as disk could then pollute the central white dwarf's atmosphere to
create a DAZ, and be observable via its IR excess. -- While the search for
planets around single white dwarfs has not been successful yet, the
existence of dusk disks has indeed been established observationally.
\par
In addition to orbital effects, \citet{2007ApJ...661.1192V} consider
the thermal conditions during the planetary nebula phase and establish
new regions in the orbit and mass parameter space where planets
survive all the way to the white dwarf stage of their host
stars. Focusing more on the orbital evolution,
\citet{2008inbook.0908.3328H} investigates the issue of survival for
planets in binary systems.  A consequence in all of these studies is
that finding planets at parameter combinations that correspond to
previously ''forbidden regions'' immediately requires that second
generation planet formation scenarios must have been at work. This can
in principle include migration and significant accretion of previously
existing planets in newly formed disks.
\par
Of course, these investigations are predominantly concerned with the
more violent effects during the AGB phase. However, once a planet has
made it safely into the orbit of a subdwarf B host star, it can be
expected to continue to exist without too much hassle when the host
evolves to its final white dwarf stage.
\par 
A first scenario that explained the case of \mbox{\object{V391 Pegasi
    b}} basically in the context of single star evolution assumes that
the planet is an old first generation planet which survived a common
envelope phase.  The situation where the star at maximum red giant
expansion almost engulfed the planet is depicted in
Fig.~\ref{fig:hs2201}. Assuming this scenario, the existence of
\object{V391 Pegasi b} proves observationally that gas-giant planets
can survive the first red giant expansion in orbits similar to that of
our Earth.
\par
\citet{2008ASPC..392..215S} puts forward alternative scenarios.  In a
second variant, \object{V391 Pegasi b} would have been an outer
planet, never really under a strong direct influence of the expanding
host star.  The exceptional mass loss of the host star, turning it
into a sdB, would instead have been triggered by another closer-in
planet that got potentially destroyed in the process.
\par
Scenario three explains \object{V391 Pegasi b} as a young second 
generation planet formed in a disk resulting from the merger of 
two \element{He} white dwarfs. The planet would then be a 
helium planet (compare also \citealt{2005ApJ...632L..37L}).
\par
In the case of the planets around HW Vir-like systems, the situation
is somewhat easier as there is no dispute about the origin of the
enhanced mass loss in a common envelope ejection phase. Besides the
argument by \citet{2009AJ....137.3181L} that the planets were formed
in a circumbinary disk, this leaves room for an interesting variety of
possible planet formation scenarios, including a second generation
scenario (cited from \citealt{2009ARA&A..47..211H}):
\textsf{\footnotesize ''At birth, \object{HW Vir}'s binary components must have been much further
apart than they are today. During this [red giant] mass-loss episode
the planets as well as the stellar companion may have gained
mass. Rauch (2000) suggested that the low-mass companion in \object{AA Dor} may
have grown from an initial planet by mass accretion during the CE-ejection 
phase. Could that have happened to the \object{HW Vir} system? As a
matter of pure speculation, the star may have been born with three
massive planets: the innermost launched a CE ejection, spiraled in,
and accreted so much material as to turn into a low-mass
star. [\ldots] Another speculation concerns in situ
formation; that is, could the planet have been formed during the CE
phase?''\normalsize}
\subsection{The fate of evolving stars with planets}
\label{subsec:fateofstars}
As already implied above, it has recently been suggested that the
presence of planets may have implications for the formation of sdB
stars. This marks the renaissance of an older idea:
\citet{1998AJ....116.1308S} suggested that planets may constitute the
\emph{second parameter} influencing the irregular morphology of the
(blue part of the) horizontal branch in globular clusters and elliptical galaxies.
The physical processes discussed as candidates for the second
parameter in horizontal branch morphology are
\begin{asparaitem}
\setlength{\itemsep}{-1.5pt}
\item age of the globular cluster,
\item deep helium mixing and radiative levitation,
\item large \element{He} abundance or fast rotation,
\item stellar density in the cluster,
\item planets enhancing mass loss on the red giant branch.
\end{asparaitem}
Quite similarly to this last suggestion \citep{1998AJ....116.1308S},
\citet{2010ApSS.S} points out that in addition to strengthening the
blue component of the normal horizontal branch, enhanced mass loss due
to planets could also be decisive in forming sub\-dwarf B stars.  The
initial prediction -- that the engulfed\linebreak planet that helps to
shed the envelope will in most cases be destroyed in the process --
remains valid, so this hypothesis is challenging to test. 
\par
However, \object{V391 Pegasi}, while most probably not capable to have
caused the enhanced mass loss of its host star itself, may well be
regarded as a possible tracer for (former?) inner planets that could
indeed have been able to cause the amount of mass loss required
(compare the second scenario in the previous section). The scenario as
a general explanation for the formation of a significant part of the
single sdB star population may have a number of unsolved problems,
mostly related to our poor understanding of the common envelope
ejection mechanism. But, independently of the theory, \mbox{EXOTIME}
and other programs now start to built up an empirical data base of
low-mass companion statistics that further discussions can be based
on. This should help to clarify the role of an -- to date perhaps
mostly still undiscovered -- population of very low-mass companions to
apparently single subdwarf B stars in the formation of these objects.
\par
Should it one day become possible to determine the\linebreak
composition of the low-mass objects found, and should\linebreak
these turn out to be planets that were most likely formed in the
helium disk of a \element{He} white dwarf merger event, the origin of
the host stars will also have been cleared up unambiguously.
\section{Summary}
\label{sec:summary}
This article has attempted to highlight the added value
of incorporating stellar pulsations in the comprehensive
investigation of various stellar and planetary systems.
\par
The potential of links between asteroseismology and\linebreak
exoplanet science was first illustrated for solar-like
oscillators. Accurate asteroseismic radii for (solar-like) host stars
of transiting exoplanets translate into good radius determinations for
the transit planets.
\par
In the context of connections with coherent pulsations, as found in
evolved stars, the focus was directed towards the investigation of
stellar and planetary systems at late stages of stellar evolution.
The asteroseismology-exoplanet-con\-nection was explored in particular
for post-RGB stars. The exotic class of subdwarf B stars, of which
many show pulsations, was introduced as a group with several new
planet discoveries. In discussing the fate of planetary systems based
on known planets around extreme horizontal branch stars
(e.g. \object{V391 Pegasi}), the question arises whether these are
first or second generation planets. Data from programs such as
\mbox{EXOTIME}, which make use of the timing method to find more such
systems, can fill a gap with respect to the sensitivity to planets in
wide orbits and applicability to evolved stars.
\par
With respect to the puzzle of subdwarf B formation, planets were
proposed to (partly) be at the origin of sub\-dwarf B stars, and also to
play a role as the second parameter for the horizontal branch
morphology in globular clusters and elliptical galaxies.
\par
The aspects this article summarizes are all from the
pre-\textit{Kepler} era. It will be fascinating to see how our 
understanding in a great many areas evolves with the new data.
\acknowledgements
The author would like to thank the Astrono\-mische Gesellschaft 
for the opportunity to present this work during the 2009
Ludwig Biermann lecture.
Warm thanks go to all of my collaborators and students who have been
part of these and many other research projects, including K.\ Werner,
S.\ Dreizler and his group, R.\ Silvotti, R.\ Lutz, R.\ Kruspe and B.\ L\"optien.
\par
Out of the various sources for data, I want to single
out the observers or operators at the Calar Alto, MONET, HET, and Keck
telescopes (special thanks to A.~Reiners and G.~Basri for providing
the \object{V391 Pegasi} spectrum), and also thank U.\ Heber and
T.\ Rauch for providing grids of model spectra.  Spectral energy
distributions (SEDs) of the hydrogen NLTE models that were calculated
with \emph{TMAP}, the T\"ubingen Model-Atmosphere Package, were
retrieved via
\emph{TheoSSA}\footnote{http://vo.ari.uni-heidelberg.de/ssatr-0.01/TrSpectra.jsp?},
a service provided by the German Astrophysical Virtual Observatory
(\emph{GAVO}\footnote{http://www.g-vo.org}).
\par
The preparation of this article has benefited from up-to-date information on systems
and WWW links of the Extra-Solar Planets Encyclopaedia maintained by J.\ Schneider
(www.exoplanet.eu).
%
%


\end{document}